\documentclass[11pt]{article} 
\usepackage{epsf}

\newcommand{\bea}{\begin{eqnarray}}
\newcommand{\eea}{\end{eqnarray}}
\newcommand{\be}{\begin{equation}}
\newcommand{\ee}{\end{equation}}

\title{Bose-Einstein condensates in symmetry breaking states}
\author{Yvan Castin and Christopher Herzog}
\date{Laboratoire Kastler-Brossel de l'\'Ecole normale sup\'erieure
\thanks{Laboratoire Kastler-Brossel is a unit\'e de recherche de l'\'Ecole
normale sup\'erieure et de l'Universit\'e Pierre et Marie Curie,
associ\'ee au CNRS.} \\
24 rue Lhomond \\ 75 231 Paris Cedex 5 \\ France}

\begin{document}
\maketitle
\begin{abstract}
We consider two models of interacting Bose gases: a gas of spin
one particles in the ground state of a cubic box 
and a one-dimension Bose gas with contact interactions. We show how to calculate
{\sl exact} eigenstates of the corresponding $N-$body Hamiltonians.
Both models
share the property of not leading to the formation of a Bose-Einstein condensate,
even at zero temperature, in the strict sense of the existence of a single
one-particle state with a macroscopic population. We show that a lot of physical 
insight can be gained on these two model systems by using the usual Hartree-Fock
mean field approach: in this approximation, that we test against the exact result,
everything happens as if a single realization of the system was a Bose-Einstein condensate
in a state $\phi$ breaking the rotational or translational symmetry,
and varying in a random way for any new experimental realization.
\end{abstract}

Since the observation in 1995 of the first Bose-Einstein condensates
in atomic gases \cite{JILA_first,MIT_first,Hulet} 
a renewed interest is taking place
for these macroscopic quantum states of matter \cite{StringariRev}.

In particular some experiments are the realization of
what was considered before 1995 as gedanken experiments. From this
point of view the interference experiment performed at MIT by the group
of Wolfgang Ketterle is particularly illustrative: two condensates
that \lq had never seen each other' are made to spatially overlap, 
leading to the formation of interference fringes \cite{MIT_franges}.
In this last sentence we have deliberately paraphrased 
P.W.\ Anderson, who raised the question 
whether two superfluids that \lq have never
seen each other' have or not a well-defined relative phase \cite{Anderson}.

The MIT-type phase experiment and more generally the concept
of phase of a matter wave field has been the subject of several
theoretical analyses \cite{Leggett,Javanainen,mesure_franges1,mesure_franges2}.
One of the theoretical issues was to reconcile two different points of
view. In the first point of view, the two condensates in the MIT experiment
are described by a Fock state, that is by a state with a well-defined
number of particles in each condensate mode; this Fock state
leads to interference fringes in a measurement of the positions
of all the atoms, as shown numerically in \cite{Javanainen}
and analytically in \cite{mesure_franges2}, but the result is not
obvious. In the second point of view a symmetry breaking
description is adopted, attributing a well-defined relative phase $\phi$
to the two condensates in a given experimental realization, 
with $\phi$ varying in an unpredictable manner for any new realization
of the experiment, with a uniform probability distribution
in $[0,2\pi[$. The symmetry breaking point of view immediately
predicts the formation of interference fringes, in a very intuitive way,
but requires
some justification, as a symmetry of the system is broken!
We refer the reader to \cite{mesure_franges2,les_houches} for a comparison
of the two points of view, the main result being that their predictions
for the interference fringes coincide in the limit of large number of particles,
and that symmetry breaking is a convenient but
by no means necessary procedure.

The goal of the present text is to illustrate other uses of symmetry
breaking, allowing one to extend the applicability of mean-field approximations
to situations where there is no condensate in the strict sense,
that is in the sense of Penrose and Onsager \cite{Penrose},
but where a so-called fragmented condensate \cite{Nozieres} is present
(section \ref{Cap:context}).
The symmetry-breaking point of view 
that we put forward here is of a physically different
nature than the $U(1)$ phase symmetry breaking description of the
above mentioned MIT interference experiment. We consider in section
\ref{Cap:spinor}
the case of so-called \lq spinor condensates', more specifically
condensates of particles of spin one, where a $SO(3)$
rotational symmetry breaking description is applied.  We treat in section
\ref{Cap:soliton} the case of one-dimension attractive Bose gases,
where a spatial translational symmetry breaking is applied to
obtain solitonic condensates.

In both cases of spinor and solitonic condensates
the procedure followed in this paper
is the same one: the ground state of the system is
calculated exactly for an appropriate model of the interaction
potential and it is of course symmetric,
while its mean-field approximation by Hartree-Fock states breaks the symmetry.
In both cases we will consider Gedanken experiments whose single outcomes
can be predicted easily from the Hartree-Fock state and with more effort
from the exact ground state, the two predictions coinciding
for a large number of particles.
This will illustrate the ability of the mean-field approximation
to allow physical predictions in an easy and transparent way,
correct in the limit of a macroscopic system.

\section{Physical context and motivation}
\label{Cap:context}
\subsection{Usual definition of a condensate}

We consider a gas of $N$ interacting indistinguishable bosonic
particles. The total
number of particles is conserved by the Hamiltonian and
the $N-$body state of the gas is defined by the 
$N-$body density operator $\hat{\sigma}_{1,\ldots,N}$, supposed to be known.
When does one say that the gas is in a Bose-Einstein condensed state ?
We answer this question in the way
formalized by Penrose and Onsager \cite{Penrose}.

One first defines the one-body density operator $\hat{\rho}_1$ as
the trace of $\hat{\sigma}_{1,\ldots,N}$ over the state of
the particles $2,\ldots,N$:
\be
\hat{\rho}_1 \equiv N \mbox{Tr}_{2,\ldots,N}\left[
\hat{\sigma}_{1,\ldots,N}\right].
\label{eq:rho1}
\ee
Note the factor $N$ in this definition, so that $\langle \vec{r}\,|
\hat{\rho}_1|\vec{r}\,\rangle$ is exactly the mean spatial density
in $\vec{r}$. The one-body density operator can also be defined
by its matrix element in any single-particle orthonormal basis
$\psi_\alpha$ using second quantization:
\be
\langle \psi_{\alpha} |\hat{\rho}_1 |\psi_{\beta}\rangle =
\langle \hat{a}_\beta^\dagger \hat{a}_\alpha\rangle
\ee
where $\hat{a}_{\alpha,\beta}$ annihilates a particle in
the state $\psi_{\alpha,\beta}$ and $\langle \ldots \rangle$
stands for the expectation value in the density operator $\hat{\sigma}_{1,\ldots,N}$.

Second one diagonalizes the hermitian operator $\hat{\rho}_1$:
\be
\hat{\rho}_1 = \sum_k n_k |\phi_k\rangle \langle\phi_k|
\ee
where the $\phi_k$'s form the orthonormal set of eigenvectors of $\hat{\rho}_1$
with eigenvalues $n_k$. All eigenvalues $n_k$ are greater than or equal to zero
and their sum is equal to $N$. One then says that a condensate is present
if there exists a level index $k$, taken by convention to be $k=0$, such that
the two following conditions are satisfied:
\bea
n_0 &\gg& n_k \quad \mbox{for all}\quad k\neq 0 \label{eq:Cond1} \\
n_0 &\sim& N \label{eq:Cond2}
\eea
Physically this means that the maximally occupied single particle
level $\phi_0$ has a mean occupation number $n_0$ much larger than any
orthogonal single particle state (according to condition (\ref{eq:Cond1}))
and that $n_0$ is macroscopic
(according to condition (\ref{eq:Cond2})), say at least on the order of 10\% of the total
number of particles in the system.

In this case $\phi_0(\vec{r}\,)=\langle\vec{r}\,|\phi_0\rangle$ 
is the condensate wavefunction, normalized to unity,
and $n_0$ is the mean number of particles in the condensate, usually
noted as $N_0$.

\subsection{Situations considered in this paper}\label{subsec:class}
We consider systems such that even at zero temperature, that
is even in the ground state, one of the two conditions
(\ref{eq:Cond1},\ref{eq:Cond2}) is violated, so that there
is no Bose-Einstein condensate in the strict sense of
\cite{Penrose}. 

More
specifically we will exemplify a class I of such systems corresponding
to a $\hat{\rho}_1$ having several eigenstates with a macroscopic
population, and a class II where none of the eigenstates of
$\hat{\rho}_1$ has a macroscopic population.

Our example for class I is a model for a gas of interacting spin one
particles, all in the ground state of a box with periodic boundary
conditions but with free spin variables. 
One then finds that because of the $SO(3)$ rotational symmetry 
of the Hamiltonian the one-body density operator $\hat{\rho}_1$ contains
three eigenstates with the same macroscopic population (equal to $N/3$):
condition (\ref{eq:Cond1}) is violated.

Our example for class II is a
one-dimension Bose gas
with attractive contact interactions in free space (that is in the absence of
confining external potential). One then finds that because of spatial
translational symmetry the eigenstates of $\hat{\rho}_1$ are plane
waves. If one normalizes these plane waves in a fictitious quantization box of size $L$
one finds that all plane waves have a population tending to zero
in the limit $L\rightarrow +\infty$: condition (\ref{eq:Cond2}) is violated.

\subsection{Methodology of the symmetry breaking point of view}

In the usual case of a Bose-Einstein condensed system in the Penrose and Onsager
sense, $\hat{\rho}_1$ has a single macroscopic eigenvalue $N_0$ with a corresponding
normalized eigenvector $|\phi_0\rangle$. When the gas is dilute, that is when
the mean interparticle separation is much larger than the 
scattering length $a$ of the interaction potential,
the fraction of particles out of the condensate $1-N_0/N$ is very small at
zero temperature and the interaction potential
can be replaced by an effective low energy contact interaction \cite{StringariRev,les_houches}.
One can then determine an approximation to $\phi$ by using the following Hartree-Fock
\lq\lq pure condensate" variational ansatz for the ground state of the gas:
\be
|\Psi\rangle^{\mbox{\scriptsize hf}} = |N:\phi\rangle,
\ee
that is a Fock state with all the $N$ particles in the same mode $\phi$. One determines
$\phi$ by minimizing the mean energy of $|\Psi\rangle^{\mbox{\scriptsize hf}}$. 
This leads in particular to the famous Gross-Pitaevskii equation for $\phi$, very successful
in predicting the properties of the Bose-Einstein condensates of alkali gases \cite{StringariRev}.
Note that after minimization there is still the freedom of a phase change of $\phi$
by a constant phase factor $e^{i\alpha}$; but this amounts to multiplying the $N-$body
state vector $|\Psi\rangle^{\mbox{\scriptsize hf}}$ by a global phase $e^{iN\alpha}$
of no physical consequence.

We shall use exactly the same Hartree-Fock variational procedure in 
our examples of a spin one Bose gas
and of a one-dimension Bose gas with attractive interactions.
Something peculiar will happen then: one finds an infinite number of minimal
energy solutions, rather than a single one (up to a global phase factor) in the usual case.
These minimal energy solutions have all exactly the same energy, and are parameterized
by a set of continuous parameter(s) $\Omega$. We note $\phi_\Omega$
the minimal energy solution corresponding to the parameter(s) $\Omega$. 
In the case of spin one particles $\Omega$ is
actually a particular direction in space in which the spin of $\phi_\Omega$ points. In the 
one-dimension Bose gas $\Omega$ gives the position of the center of mass associated
to $\phi_\Omega$.

This structure of the results is a direct consequence of the fact that minimal
energy solutions of the Hartree-Fock ansatz are not symmetric with respect to
the symmetry group of the Hamiltonian, as they privilege some arbitrary spin direction
(for the spin one gas) or are a localized soliton around some arbitrary point of space
(for the one-dimension Bose gas). Each $\phi_\Omega$ is said to break the symmetry, 
the rotational $SO(3)$ symmetry for the spins or the translational symmetry for the 1D
Bose gas. From any given $\phi_{\Omega_0}$ one then obtains a continuous family
of $\phi_\Omega$ with the same mean energy by applying the continuous symmetry group of
the Hamiltonian to $\phi_{\Omega_0}$, that is arbitrary spin rotations or arbitrary spatial translations.

How can we then construct a reasonable approximation to the exact ground state $|\Psi_0\rangle$ of
the gas from this continuous family of Hartree-Fock states? The simplest choice
is the so-called symmetry-breaking prescription, approximating the pure state
$|\Psi_0\rangle$ by the following $N-$body density operator:
\be
\hat{\sigma}_{1,\ldots,N}^{\mbox{\scriptsize hf}}
=\int \frac{d\Omega}{\cal V} |N:\phi_\Omega\rangle\langle N:\phi_\Omega|,
\label{eq:do}
\ee
that is by a statistical mixture of all the minimal energy Hartree-Fock ansatz.
The normalization factor $\cal V$ is independent of $\Omega$ for
a convenient choice of the parameterization of the symmetry group \cite{Haar}: 
e.g.\ for $SO(3)$ one integrates over all possible solid angles $\Omega$. 
Why is the statistical mixture (\ref{eq:do}) called a symmetry breaking 
description, while it is obviously invariant
by the symmetry group ? Well, this prescription
allows one to {\sl imagine} that each particular experimental
realization of the gas corresponds to a condensate with all particles in the same state
$\phi_\Omega$, where $\Omega$ varies in a random, unpredictable way for any
new realization of the experiment: each particular experimental realization therefore
\lq breaks' the symmetry.
This point of view should of course not be taken 
too literally: it is only a convenient reinterpretation of the density operator
(\ref{eq:do}). The same density operator can also be obtained as a statistical
mixture of other $N-$body states than the $\phi_\Omega$ 
(e.g.\ a statistical mixture of its eigenstates) which may lead to a very different
physical picture. What matters actually is that
the quantum mechanical prediction of the results of any measurement performed
on the gas will be expressed as the expectation value of some operator $\hat{O}$
in the density operator $\hat{\sigma}_{1,\ldots,N}$
\cite{mesure_franges2,les_houches}:
\be
\langle \hat{O}\rangle = \mbox{Tr}\left[\hat{O}\hat{\sigma}_{1,\ldots,N}\right]
\ee
so that physical
predictions depend only on $\hat{\sigma}_{1,\ldots,N}$ as a whole, not on a particular
decomposition of $\hat{\sigma}_{1,\ldots,N}$ as a statistical mixture. For example
the probability density of measuring the particle 1 in point of real space $\vec{r}_1$,
$\ldots$, the particle $N$ in point $\vec{r}_N$ corresponds to the expectation value
of the $N-$body
observable $\hat{O}=
\hat{\Psi}^\dagger(\vec{r}_N) \ldots \hat{\Psi}^\dagger(\vec{r}_1)
\hat{\Psi}(\vec{r}_1) \ldots \hat{\Psi}(\vec{r}_N)/N!$, where $\hat{\Psi}$
is the atomic field operator \cite{Javanainen}.

There exists a more elaborate way to approximate the 
$N-$body ground state of the gas by a pure state constructed
from the Hartree-Fock states $|N:\phi_\Omega\rangle$.
The idea is to form a linear combination of these Hartree-Fock states having the
same symmetry as the exact ground state. This is achieved with the
use of a projection theorem associated to the representations
in the Hilbert space of the symmetry group of the Hamiltonian \cite{Wu}.
For a continuous symmetry group
this amounts to performing integrals of the form
\be
|\Psi_0\rangle^{\rm hfs} =  \int d\Omega f(\Omega)|N:\phi_\Omega\rangle
\label{eq:ps}
\ee
for an appropriate choice of the weight factor $f(\Omega)$. We restrict 
for simplicity to the case of a non-degenerate
ground state, associated to a unidimensional irreducible representation
of the symmetry group; the weight factor $f(\Omega)$ then has a modulus independent of $\Omega$.
While mathematically more satisfactory than (\ref{eq:do}) this form 
is not particularly illuminating from a physical point of view.
It is also more difficult to manipulate. Let us indeed try to calculate 
the expectation value of an arbitrary $k$--body operator $\hat{O}$ 
in the state (\ref{eq:ps}). In first quantized form $\hat{O}$ is a 
sum over all possible $k$--tuplets of particles of an operator of $k$ particles:
\be
\hat{O} = \sum_{1\leq i_1 < \ldots < i_k \leq N} O(i_1,\ldots, i_k).
\ee
Using the bosonic symmetry of the Hartree-Fock states one finds that all the 
$\left(\begin{array}{c}N\\ k\end{array} \right)=N!/[k!(N-k)!]$ 
possible $k$-uplets
give the same contribution to the expectation value so that
\be
{}^{\rm hfs}\langle \Psi_0|\hat{O} |\Psi_0\rangle^{\rm hfs}
= \left(\begin{array}{c}N\\ k\end{array} \right) 
\int d\Omega f(\Omega) \int d\Omega' f^*(\Omega')
\frac{\langle k:\phi_{\Omega'}|O(1,\ldots,k)|k:\phi_{\Omega}\rangle}
{\left( \langle\phi_{\Omega'}|\phi_{\Omega}\rangle \right)^k}
\left( \langle\phi_{\Omega'}|\phi_{\Omega}\rangle \right)^{N}.
\label{eq:ev}
\ee
In general the double integral over $\Omega$
and $\Omega'$ in Eq.(\ref{eq:ev}) is  difficult to calculate. 
In the case $k\ll N$, one can fortunately use a large $N$ expansion as follows.
The scalar product of $\phi_{\Omega}$ and $\phi_{\Omega'}$
raised to the very large power $N$ is a much narrower function of the distance
$\Omega-\Omega'$ than the remaining part of the integrand so that it can be approximated
by a $\delta$ distribution \cite{mesure_franges2}:
\be
\left( \langle\phi_{\Omega'}|\phi_{\Omega}\rangle \right)^{N}
\simeq  {\cal N} \delta(\Omega-\Omega')
\ee
where $\cal N$ is a constant factor. The expectation value (\ref{eq:ev}) then reduces 
to
\be
{}^{\rm hfs}\langle \Psi_0|\hat{O} |\Psi_0\rangle^{\rm hfs} \simeq
\left(\begin{array}{c}N\\ k\end{array} \right) {\cal N}
\int d\Omega |f(\Omega)|^2 \langle k:\phi_{\Omega}|O(1,\ldots,k)|k:\phi_{\Omega}\rangle.
\label{eq:reduction}
\ee
As $|f|$ has a constant modulus, this expression is equivalent to
the symmetry-breaking prediction (\ref{eq:do})!

One could then go to next order in the $1/N$ expansion, to identify the differences in
the predictions of (\ref{eq:do}) and (\ref{eq:ps}).
In general and as we shall see
for the one-dimension Bose gas, this is not worth the effort
as both ${}^{\rm hfs}\langle \Psi_0|\hat{O} |\Psi_0\rangle^{\rm hfs}$ and the symmetry-breaking
prediction Tr$[\hat{O}\sigma_{1,\ldots,N}^{\mbox{\scriptsize hf}}]$ differ from the exact
result by a term of the same order of magnitude, of order $1/N$ times the exact
result. An amusing exception is our model
of spin one particles where (\ref{eq:ps}) actually coincides with the exact ground spin state!

\section{The ground state of spinor condensates}
\label{Cap:spinor}
The alkali atoms used in the Bose-Einstein condensation experiments have an
hyperfine structure in the ground state, each hyperfine level having several Zeeman
sublevels.
Consider for example ${}^{23}$Na atoms used at MIT in the group of Wolfgang Ketterle.
The ground state has an hyperfine splitting between the lower multiplicity 
of angular momentum $F=1$ and the higher multiplicity of angular momentum $F=2$.
All the three Zeeman sublevels $m_F =0,\pm 1$ of the lower multiplicity $F=1$
cannot be trapped in a magnetic
trap (if $m_F=-1$ is trapped then $m_F=+1$ which experiences an opposite Zeeman
shift is antitrapped). But they can all be trapped in an optical dipole trap,
produced with a far off-resonance laser beam, as
the Zeeman sublevels experience then all the same lightshift.
This optical trapping was performed at MIT \cite{MIT_optical_trap}, opening the way to a series of
interesting experiments with condensates of particles 
of spin one \cite{MIT_spin1}.

We introduce first a model interaction potential for spin one particles.
We then concentrate on a specific aspect, the ground state of the spinor part of the problem,
assuming that 
the atoms are all in the ground state of a cubic box with periodic
boundary conditions.

\subsection{A model interaction potential}
In the theory of Bose-Einstein condensates the interaction potential between
the particles is usually replaced by a model interaction potential having the 
same collisional properties at low energy as the \lq exact' atomic interaction
potential. This model interaction potential is conveniently taken to be Fermi's pseudo-potential
initially introduced in nuclear physics \cite{Fermi,Huang,Jean_varenna,les_houches}.
It has the following action on a two-body wavefunction $\psi_{1,2}$:
\be
\langle\vec{r}_1,\vec{r}_2\,|V(1,2)|\psi_{1,2}\rangle 
= g \delta(\vec{r_1}-\vec{r_2})   \left[{\partial\over\partial{r_{12}}}\left(r_{12}
\ \psi_{1,2}(\vec{r}_1,\vec{r}_2) \ \right)\right].
\label{eq:ppf}
\ee
This is essentially a contact interaction potential, conveniently regularized,
and with a coupling constant
$g$ proportional to the $s$-wave scattering length $a$a of the \lq exact' interaction potential:
\be
g=\frac{4\pi\hbar^2}{m} a.
\ee

We have to generalize the model 
scalar pseudo-potential of Eq.(\ref{eq:ppf}) to the case of particles having
a spin different from zero. As we want to keep the simplicity of a contact
interaction potential we choose the simple form
\be
V(1,2) \equiv
{\cal V}_{\mbox{\scriptsize spin}}(1,2)
\delta(\vec{r_1}-\vec{r_2}) \left[{\partial\over\partial{r_{12}}}\left(r_{12}
\ \cdot \ \right)\right]
\ee
that is the product of an operator acting only on the spin of the particles 1 and 2,
and of the usual regularized contact interaction acting only on the relative motion
of the two particles. 
The interaction potential $V(1,2)$ has to be invariant 
by a simultaneous rotation of the spin variables and of the position variables 
of the two particles. As the contact interaction is already rotationally invariant,
the spin part of the interaction ${\cal V}_{\mbox{\scriptsize spin}}(1,2)$ has to
be invariant by any simultaneous rotation of the two spins. 

This condition of rotational invariance of ${\cal V}_{\mbox{\scriptsize spin}}(1,2)$
is easy to express in the coupled basis obtained by the composition of the two spins
of particle 1 and particle 2: within each subspace of well defined total angular
momentum ${\cal V}_{\mbox{\scriptsize spin}}(1,2)$ has to be a scalar. Let us restrict
to the case studied at MIT, 
with spin one particles. By composition of $F=1$ and $F=1$
we obtain a total angular momentum $F_{\mbox{\scriptsize tot}}=2$, 1 or 0, so that
one can write
\be
{\cal V}_{\mbox{\scriptsize spin}}(1,2)= g_2 P_{F_{\mbox{\scriptsize tot}}=2}(1,2)
+g_1 P_{F_{\mbox{\scriptsize tot}}=1}(1,2) + g_0 P_{F_{\mbox{\scriptsize tot}}=0}(1,2)
\ee
where the $g$'s are coupling constants and
the $P(1,2)$'s are projectors on the subspace of particles 1 and 2 with a well
defined total angular momentum $F_{\mbox{\scriptsize tot}}$.
At this stage we can play a little trick, using the fact that the states of
$F_{\mbox{\scriptsize tot}}=1$ are antisymmetric by the exchange of particles 1 and 2 (whereas
the other subspaces are symmetric). The regularized contact interaction scatters
only in the $s$-wave, where the external wavefunction of atoms 1 and 2
is even by the exchange of the positions $\vec{r}_1$ and $\vec{r}_2$; as our atoms are
bosons, the spin part has also to be symmetric by exchange of the spins of atoms
1 and 2 so that the \lq fermionic' part of ${\cal V}_{\mbox{\scriptsize spin}}(1,2)$, that is
in the subspace $F_{\mbox{\scriptsize tot}}=1$, has no effect. We can therefore change
$g_1$ at will without affecting the interactions between bosons. The most convenient
choice is to set $g_1=g_2$ so that we obtain
\be
{\cal V}_{\mbox{\scriptsize spin}}(1,2)= g_2 \mbox{Id}(1,2) 
+(g_0-g_2) P_{F_{\mbox{\scriptsize tot}}=0}(1,2) 
\label{eq:Vs}
\ee
where $\mbox{Id}$ is the identity.
The subspace $F_{\mbox{\scriptsize tot}}=0$ is actually of dimension one, and it
is spanned by the vanishing total angular momentum state $|\psi_0(1,2)\rangle$. Using the
standard basis $|m=-1,0,+1\rangle$ of single particle
angular momentum with $z$ as quantization axis, one can write
\be
|\psi_0(1,2)\rangle = -\frac{1}{\sqrt{3}}\left[ |+1,-1\rangle+|-1,+1\rangle -|0,0\rangle\right].
\label{eq:spin0_s}
\ee
A more symmetric writing is obtained in the single particle angular momentum
basis $|x,y,z\rangle$ used in chemistry,  defined by
\bea
|+1\rangle &=& -\frac{1}{\sqrt{2}}\left( |x\rangle + i|y\rangle\right) \\
|-1\rangle &=& +\frac{1}{\sqrt{2}}\left( |x\rangle - i|y\rangle\right) \\
|0\rangle &=& |z\rangle.
\eea
The vector $|\alpha\rangle$ in this basis $(\alpha=x,y,z)$
is an eigenvector of angular momentum along axis
$\alpha$ with the eigenvalue zero. One then obtains
\be
|\psi_0(1,2)\rangle = \frac{1}{\sqrt{3}} \left[ |x,x\rangle +|y,y\rangle + |z,z\rangle
\right].
\label{eq:spin0_c}
\ee

To summarize, the part of the Hamiltonian describing the interactions between
the particles can be written, if one forgets for simplicity the regularizing
operator in the pseudo-potential:
\bea
H_{\mbox{\scriptsize int}} &=& \frac{g_2}{2} \int d^3\vec{r}\,
\sum_{\alpha,\beta=x,y,z} \hat{\psi}_\alpha^\dagger \hat{\psi}_\beta^\dagger
\hat{\psi}_\beta \hat{\psi}_\alpha  \nonumber \\
&& + \frac{g_0-g_2}{6}\int d^3\vec{r}\, \sum_{\alpha,\beta=x,y,z}
\hat{\psi}_\alpha^\dagger\hat{\psi}_\alpha^\dagger
\hat{\psi}_\beta\hat{\psi}_\beta
\label{eq:model_hamil}
\eea
where $\hat{\psi}_{\alpha}(\vec{r}\,)$ is the atomic field operator
for the spin state $|\alpha\rangle$.
This model Hamiltonian has also been proposed by \cite{Ho,Zhang,Machida}.

We now restrict to the spinor part of the $N-$body problem, assuming
that all the $N$ particles of the gas are Bose-Einstein condensed
in the ground state of a cubic box of size $L$ with periodic boundary 
conditions. This is a reasonable assumption for a dilute Bose gas
with effective repulsive interactions ($g_2,g_0\geq 0$). It amounts
to approximating the atomic field operators by
\be
\hat{\psi}_{\alpha}(\vec{r}\,) \simeq \frac{1}{L^{3/2}} \hat{a}_\alpha
\ee
where the operator $\hat{a}_\alpha$ annihilates a particle in the state
$|\vec{k}=\vec{0}\rangle|\alpha\rangle$ ($\alpha=x,y,z$) that is in the
plane wave of vanishing wavevector in the box and with internal spin 
in state $|\alpha\rangle$. This leads to our final model Hamiltonian:
\be
H_{\mbox{\scriptsize spin}} = \frac{g_2}{2L^3} \sum_{\alpha,\beta=x,y,z} 
\hat{a}_\alpha^\dagger \hat{a}_\beta^\dagger \hat{a}_\beta \hat{a}_\alpha 
+\frac{1}{6L^3}(g_0-g_2)\hat{A}^\dagger \hat{A}
\label{eq:hspin}
\ee
where we have introduced
\be
\label{eq:definit_A}
\hat{A} = \hat{a}_x^2+\hat{a}_y^2+\hat{a}_z^2.
\ee
Up to a numerical factor 
$\hat{A}$ annihilates a pair of particles in the two-particle spin state 
$|\psi_0(1,2)\rangle$
of vanishing total angular momentum, as shown by Eq.(\ref{eq:spin0_c}).
Note that we suppose that there is no magnetic field
applied to the sample; the effect of a magnetic field is considered
in \cite{Ho_preprint}.

\subsection{Ground state in the Hartree-Fock approximation}

We now minimize the energy of the $N$ spins within the Hartree-Fock trial 
state vectors $|N:\phi\rangle$ with the constraint that
$|\phi\rangle$ is normalized to unity
but without any constraint on the total angular momentum of the spins.
The external part of the condensate wavefunction is simply the plane wave with momentum
$\vec{k}=\vec{0}$ whereas the spinor part of the wavefunction remains
to be determined:
\be
\langle \vec{r}\,|\phi\rangle = \frac{1}{L^{3/2}}\sum_{\alpha=x,y,z} c_\alpha |\alpha\rangle
\ \ \ \mbox{with}\ \ \ \sum_\alpha |c_\alpha|^2=1.
\label{eq:cond_norm_spin}
\ee
From the model interaction Hamiltonian (\ref{eq:model_hamil}) 
the mean energy per particle in a Hartree-Fock state is easy to find,
using the identity making Hartree-Fock states so convenient:
\bea
\hat{\psi}_{\alpha}(\vec{r}\,) |N:\phi\rangle &=& N^{1/2} \langle \vec{r},\alpha|\phi\rangle
|N-1:\phi\rangle \\
&=& \left(\frac{N}{L^3}\right)^{1/2} c_\alpha |N-1:\phi\rangle.
\eea
One then finds the mean energy per particle:
\be
\frac{E}{N} = \frac{N-1}{2L^3} g_2 +\frac{N-1}{6L^3} (g_0-g_2) |A|^2
\label{eq:ener_spin}
\ee
where we have introduced the complex quantity
\be
A = \sum_{\alpha=x,y,z} c_\alpha^2=\vec{c}\,^2
\ee
where $\vec{c}$ is the vector of components $(c_x,c_y,c_z)$.
We have to minimize the mean energy over the state of the spinor.

\begin{itemize}
\item Case $g_2> g_0$
\end{itemize}
This is the case of sodium \cite{MIT_spin1}. As the coefficient $g_0-g_2$ is negative
in Eq.(\ref{eq:ener_spin}) we have to maximize the modulus of the complex quantity
$A$. As the modulus of a sum is less than the sum of the moduli we immediately
get the upper bound
\be
|A|\leq \sum_{\alpha=x,y,z} |c_\alpha|^2 =1
\ee
leading to the minimal energy per particle
\be
\frac{E}{N} = \frac{N-1}{2L^3}g_2  +\frac{N-1}{6L^3}(g_0-g_2).
\ee
The upper bound for $|A|$
is reached only if all complex numbers $c_\alpha^2$ have the same
phase modulo $2\pi$. This means that one can write 
\be
c_\alpha = e^{i\theta} n_\alpha
\label{eq:def_n}
\ee
where $\theta$ is a constant phase and $\vec{n}=(n_x,n_y,n_z)$ is any unit vector
with real components. Physically this corresponds to a spinor wavefunction
being the zero angular momentum state for a quantization axis pointing in the direction
$\vec{n}$. 
The direction $\vec{n}$ is well defined in the Hartree-Fock
ansatz, but it is arbitrary as no spin direction is privileged by the Hamiltonian.
We are facing symmetry breaking, here a rotational $SO(3)$ symmetry breaking.
The symmetry breaking prescription (\ref{eq:do}) then leads to the density
operator
\be
\hat{\sigma}_{1,\ldots,N}^{\mbox{\scriptsize hf}}
=\int \frac{d^2\vec{n}}{4\pi} |N:0_{\vec{n}}\rangle\langle N:0_{\vec{n}}|,
\label{eq:do_cas1}
\ee
where the integral is taken over the unit sphere, that is over all solid angles.
and where we have introduced the single particle state
\be
|0_{\vec{n}}\rangle = n_x|x\rangle +n_y|y\rangle + n_z|z\rangle.
\ee

\begin{itemize}
\item Case $g_2< g_0$
\end{itemize}
In this case we have to minimize $|A|$ to get the minimum of energy. The minimal value
of $|A|$ is simply zero, corresponding to spin configurations such that
\be
\vec{c}\,^2 \equiv \sum_{\alpha=x,y,z} c_\alpha^2 =0
\label{eq:carre_zero}
\ee
with an energy per particle
\be
\frac{E}{N} = \frac{N-1}{2L^3} g_2.
\ee
To get more physical understanding we split the vector $\vec{c}$ as
\be
\vec{c} = \vec{R} + i\vec{I}
\ee
where the vectors $\vec{R}$ and $\vec{I}$ have purely real components. Expressing the 
fact that the real part and imaginary part of $\vec{c}\,^2$ vanish, and 
using the normalization condition 
$\vec{c}\cdot\vec{c}\,^*=1$ in Eq.(\ref{eq:cond_norm_spin}) we finally obtain
\bea
\vec{R}\cdot\vec{I} &=& 0 \\
\vec{R}\,^2 &=& \vec{I}\,^2 = \frac{1}{2}.
\eea
This means that the complex vector $\vec{c}$ is circularly polarized with respect
to the axis $\vec{n}$ orthogonal to $\vec{I}$ and $\vec{R}$. Physically this corresponds
to a spinor condensate wavefunction having an angular momentum $\pm \hbar$ along
the axis $\vec{n}$. The direction of axis $\vec{n}$ is well defined in the Hartree-Fock ansatz
but it is arbitrary.  
The symmetry breaking prescription (\ref{eq:do}) then leads to the density
operator
\be
\hat{\sigma}_{1,\ldots,N}^{\mbox{\scriptsize hf}}
=\int \frac{d^2\vec{n}}{4\pi} |N:+_{\vec{n}}\rangle\langle N:+_{\vec{n}}|
\label{eq:do_cas2}
\ee
where $|+_{\vec{n}}\rangle$ is the spin state of angular momentum $+\hbar$
along axis $\vec{n}$.

\subsection{Exact ground state vs Hartree-Fock for the spinor model}
The Hamiltonian (\ref{eq:hspin}) can be diagonalized exactly \cite{edhs}. This
is not surprising as (i) it is rotationally invariant and (ii) the 
bosonic $N-$particle
states with a well defined total angular momentum $S_{N}$ can be calculated:
one finds that $S_{N}=N,N-2,\ldots$, leading to degenerate multiplicities
of $H_{\mbox{\scriptsize spin}}$ of degeneracy $2S_{N}+1$.  As a consequence the
one-body density operator $\hat{\rho}_1$ defined in (\ref{eq:rho1}) calculated
for the ground state of the gas is spin rotationally invariant; as it acts on an irreducible
representation of $SO(3)$, of spin one, it is scalar according to Schur's lemma
\cite{Wu}:
\be
\hat{\rho}_1 = \frac{N}{3} |\vec{k}=\vec{0}\rangle\langle\vec{k}=\vec{0}|
\otimes \mbox{Id}_{\mbox{\scriptsize spin}}.
\label{eq:rho_1_spin}
\ee
We are therefore in the class I of section \ref{subsec:class}.

To derive the exact spectrum of the spin Hamiltonian
one may use in practice the following tricks: The double sum proportional to
$g_2$ in Eq.(\ref{eq:hspin}) can be expressed in terms of the operator total number of
particles $\hat{N}$ only, 
\be
\hat{N}=\sum_\alpha \hat{a}_\alpha^\dagger \hat{a}_\alpha.
\ee
So diagonalizing $H_{\mbox{\scriptsize spin}}$ amounts to
diagonalizing $\hat{A}^\dagger \hat{A}$!
Second the total momentum operator $\hat{\vec{S}}$
of the $N$ spins, defined as the sum of all the spin operators of
the individual atoms in units of $\hbar$,
can be checked to satisfy the identity 
\be
\hat{\vec{S}}\cdot\hat{\vec{S}} + \hat{A}^\dagger \hat{A} = \hat{N}(\hat{N}+1) 
\label{eq:iden_avec_s}
\ee
so that the Hamiltonian for $N$ particles becomes a function of $\hat{\vec{S}}$
\cite{edhs}:
\be
H_{\mbox{\scriptsize spin}} = \frac{g_2}{2L^3} \hat{N}(\hat{N}-1)
+\frac{1}{6L^3}(g_0-g_2)\left[\hat{N}(\hat{N}+1) - \hat{\vec{S}}\cdot\hat{\vec{S}}\right].
\label{eq:hspin_simple}
\ee
We recall that $\hat{\vec{S}}\cdot\hat{\vec{S}}=S_{N}(S_{N}+1)$ within the subspace of total spin
$S_{N}$.

When $g_2<g_0$ the ground state of $H_{\mbox{\scriptsize spin}}$ corresponds to the
multiplicity $S_{N}=N$, containing {\sl e.g.} the state with all the
spins in the state $|+\rangle$. In this case the $N-$particle states obtained
with the Hartree-Fock approximation 
are exact eigenstates of $H_{\mbox{\scriptsize spin}}$, and the symmetry breaking
prescription (\ref{eq:do_cas2}) correctly gives the projector on the maximal
spin multiplicity.

When $g_2>g_0$ the ground state of $H_{\mbox{\scriptsize spin}}$ corresponds to the 
multiplicity of minimal total angular momentum, $S_{N}=1$ for $N$ odd
or $S_{N}=0$ for $N$ even. In this case the Hartree-Fock symmetry
breaking prescription (\ref{eq:do_cas1})
is only an approximation of the exact ground state of $H_{\mbox{\scriptsize spin}}$. 
The error on the energy per
particle tends to zero in the thermodynamical limit; for $N$ even one finds indeed
\be
\frac{\delta E}{N} = -\frac{1}{3L^3}(g_0-g_2).
\ee

But what happens if one uses the pure state, symmetry restored approximation
(\ref{eq:ps}) by coherently summing up the Hartree-Fock
ansatz over the direction $\vec{n}$ defined in Eq.(\ref{eq:def_n})? Assume that
$N$ is even; one has then to reconstruct from the Hartree-Fock ansatz a rotationally
invariant state. This amounts
to considering the following normalized state for the $N$ spins:
\be
|\Psi_0\rangle^{\mbox{\scriptsize hfs}} =  \sqrt{N+1}
\int \frac{d^2\vec{n}}{4\pi}\, |N:0_{\vec{n}}\rangle
\label{eq:exact}
\ee
The state vector $|\Psi_0\rangle^{\mbox{\scriptsize hfs}}$, being non zero and having a vanishing 
total angular momentum,
is equal to the exact ground state of $H_{\mbox{\scriptsize spin}}$:
\be
|\Psi_0\rangle^{\mbox{\scriptsize hfs}} = |\Psi_0\rangle .
\ee

The expression (\ref{eq:exact}) can then be used as a starting point to obtain
various forms of $|\Psi_0\rangle$. If one expresses the Hartree-Fock state as the 
$N-$th power of the creation operator $\sum_\alpha \hat{a}_{\alpha}^{\dagger} n_{\alpha}$
acting on the vacuum $|\mbox{vac}\rangle$,
and if one expands this power with the usual binomial formula, the integral
over $\vec{n}$ can be calculated explicitly term by term and one obtains:
\be
|\Psi_0\rangle = {\cal N} \left(\hat{A}^\dagger\right)^{N/2} |\mbox{vac}\rangle
\label{eq:cdp}
\ee
where $\cal N$ is a normalization factor and the operator
$\hat{A}$ is defined in Eq.(\ref{eq:definit_A}). 
Formula (\ref{eq:cdp}) indicates that $|\Psi_0\rangle$
is simply a \lq condensate' of pairs in the state $|\psi_0(1,2)\rangle$. 
It can be used
to expand $|\Psi_0\rangle$ over Fock states with a well 
defined number of particles in the
modes $m=0,m=\pm 1$, reproducing Eq.(13) of \cite{edhs}.

To be complete we mention another way of constructing the exact eigenvectors
and energy spectrum of $H_{\mbox{\scriptsize spin}}$. The idea is to diagonalize $\hat{A}^\dagger
\hat{A}$ using the fact that $\hat{A}$ obeys a commutation relation 
that is reminiscent of that
of an annihilation operator:
\be
[\hat{A},\hat{A}^\dagger ] = 4\hat{N}+6.
\ee
In this way $\hat{A}^\dagger$ acts as a raising operator: acting on an eigenstate 
of $\hat{A}^\dagger\hat{A}$ with eigenvalue $\lambda$ and $N$ particles,
it gives an eigenstate of $\hat{A}^\dagger\hat{A}$ with eigenvalue $\lambda+4 N+6$
and with $N+2$ particles. One can also check from the identity (\ref{eq:iden_avec_s})
that the action of $\hat{A}^\dagger$ does not change the total spin:
\be
[\hat{A}^\dagger, \hat{\vec{S}}\cdot \hat{\vec{S}} ]=0.
\ee
By repeated actions of $\hat{A}^\dagger$ starting from
the vacuum one arrives at Eq.(\ref{eq:cdp}), creating the eigenstates with $N$ even
and vanishing total spin $S=0$. By repeated actions of $\hat{A}^\dagger$ starting from
the eigenstates with $N=2$ and total spin $S=2$ ({\sl e.g}.\ the state $|++\rangle$) one
obtains all the states with $N$ even and total spin $S=2$. 
More generally the eigenstate of $H_{\mbox{\scriptsize spin}}$ with total spin $S$, a spin
component $m=S$ along $z$ and $N$ particles is:
\be
||N,S,m=S\rangle \propto \left(\hat{A}^\dagger\right)^{(N-S)/2} 
|S:+1\rangle
\label{eq:form_gene}
\ee
where $|S:+1\rangle$ represents $S$ particles in the state $|+1\rangle$.
From Eq.(\ref{eq:form_gene})
one can generate the states with spin components $m=S-1,\ldots,-S$ 
by repeated actions of the spin-lowering operator $\hat{S}_-=\hat{S}_x-i\hat{S}_y$
in the usual way.
We note that formula (\ref{eq:form_gene}) was derived
independently in \cite{Ho_preprint}.

\subsection{Advantage of a symmetry breaking description}\label{subsec:advan}
We now illustrate the physically transparent character of the symmetry breaking 
prescription by analyzing the following gedanken experiment.
Imagine that we have prepared a gas of sodium atoms ($g_2>g_0$)
in the collective ground spin state with an even number of particles $N$, and that we
let the atoms leak one by one out of the trap, in a way that does not
perturb their spin. We then measure the spin component along $z$ of the outgoing
atoms. Suppose that we have performed this measurement on $k$ atoms, with
$k\ll N$. We then raise
the simple question: what is the probability $p_k$ that all the $k$ detections give
a vanishing angular momentum along $z$?

Let us start with a naive reasoning based on the one-body density matrix
of the condensate.
The mean occupation numbers of the single particle spin states $|m=-1\rangle$,
$|m=0\rangle$ and $|m=+1\rangle$ in the initial condensate are all equal to $N/3$,
see Eq.(\ref{eq:rho_1_spin}).
The probability of detecting
the first leaking atom in $|m=0\rangle$ is therefore $1/3$. Naively we assume that since
$k\ll N$ the detections have a very weak effect on the state of the condensate
and the probability of detecting the $n-$th atom ($n\leq k$)
in the $m=0$ channel is nearly independent of the $n-1$ previous detection results.
The probability for $k$ detections in the $m=0$ channel should then be
\be
p_k^{\mbox{\scriptsize naive}} = \frac{1}{3^k}.
\label{eq:Naive}
\ee

Actually this naive reasoning is wrong (and by far) as soon as $k\geq 2$. The first detection
of an atom in the $m=0$ channel projects 
the spin state of the remaining atoms in
\be
|\Psi_1\rangle = {\cal N}_1 \hat{a}_0 |\Psi_0\rangle
\ee
where $\hat{a}_0$ annihilates an atom in spin state $m=0$, $|\Psi_0\rangle$ is the exact
collective spin ground state and ${\cal N}_1$ is a normalization factor.
The probability of detecting the second atom in $m=0$ (knowing that the first atom
was detected in $m=0$) is then given by
\be
\frac{p_2}{p_1} = \frac{\langle\Psi_1|\hat{a}_0^\dagger\hat{a}_0|\Psi_1\rangle}
{\langle\Psi_1|\sum_{m=-1}^{+1} \hat{a}_m^\dagger\hat{a}_m|\Psi_1\rangle}.
\ee
The denominator is simply equal to $N-1$ as $|\Psi_1\rangle$ is a state with
$N-1$ particles. Using the integral form (\ref{eq:exact}) and the simple
effect of an annihilation operator on a Hartree-Fock state, {\sl e.g.}
\be
\hat{a}_0^2|N:0_{\vec{n}}\rangle =\left[N(N-1)\right]^{1/2}n_z^2 |N-2:0_{\vec{n}}\rangle
\ee
we are able to express the probability in terms of integrals over solid angles:
\be
\frac{p_2}{p_1} = \frac{\displaystyle\int d^2\vec {n} \int d^2\vec{n}\,' \ 
n_z^2 n_z'^2(\vec{n}\cdot\vec{n}\,')^{N-2}}
{\displaystyle\int d^2\vec {n} \int d^2\vec{n}\,' \ n_z n_z'
(\vec{n}\cdot\vec{n}\,')^{N-1}}.
\ee
We suggest the following procedure to calculate these integrals. One first integrates
over $\vec{n}\,'$ for a fixed $\vec{n}$, using spherical coordinates relative to
the \lq vertical' axis directed along $\vec{n}$: the polar angle $\theta'$ is then the 
angle between $\vec{n}\,'$ and $\vec{n}$ so that one has simply
$\vec{n}\cdot\vec{n}\,'=\cos\theta'$. The integral over $\theta'$ and over the azimuthal
angle $\phi'$ can be performed, giving a result involving only $n_z$. The remaining integral
over $\vec{n}$ is performed with the spherical coordinates of vertical axis $z$.
This leads to
\be
\frac{p_2}{p_1} = \frac{3}{5} +\frac{2}{5(N-1)}.
\ee

The ratio $p_2/p_1$ is therefore different from the naive (and wrong!)
prediction (\ref{eq:Naive}). For $N=2$ one finds $p_2/p_1=1$ so that
the second atom is surely in $m=0$ if the first atom was detected in $m=0$.
As the two atoms were initially in the state with total angular momentum
zero, this result could be expected from the expression (\ref{eq:spin0_s}) of
the two-particle spin state. In the limit of large $N$ we find
that once the first atom has been detected in the $m=0$ channel, the probability
$p_2/p_1$
for detecting the second atom in the same channel $m=0$ is $3/5$. This somehow
counter-intuitive result shows that the successive detection probabilities
are strongly correlated in the case of the spin state (\ref{eq:exact}).

The exact calculation of the ratio
\be
\frac{p_{k+1}}{p_k} = \frac{\displaystyle\int d^2\vec {n} \int 
d^2\vec{n}\,' \ 
n_z^{k+1} n_z'^{k+1}(\vec{n}\cdot\vec{n}\,')^{N-(k+1)}}
{\displaystyle \int d^2\vec {n} \int d^2\vec{n}\,' \ n_z^k n_z'^k
(\vec{n}\cdot\vec{n}\,')^{N-k}}
\ee
gets more difficult when $k$ increases. The large $N$ limit
for a fixed $k$ is easier to obtain, paraphrasing the
reasoning leading to Eq.(\ref{eq:reduction}): in the integral over $\vec{n}\,'$
the function $(\vec{n}\cdot\vec{n}\,')^{N-(k+1)}$
is extremely peaked around $\vec{n}'=\vec{n}$ so that we can replace the slowly
varying function $n_z'^{k+1}$
by $n_z^{k+1}$. This leads to
\be
\lim_{N\rightarrow+\infty} \frac{p_{k+1}}{p_k} = \frac{2k+1}{2k+3}.
\label{eq:exact_ratio}
\ee

We now give the reasoning in the symmetry breaking point of view (\ref{eq:do_cas1}), which
assumes that a single experimental realization of the condensate corresponds
to a Hartree-Fock state $|N:0_{\vec{n}}\rangle$ with the direction
$\vec{n}$ being an unpredictable random variable with uniform distribution
over the sphere. If the system is initially in the spin state $|N:0_{\vec{n}}\rangle$
there is no correlation between the spins, and the probability of having $k$ detections
in the channel $m=0$ is simply $(n_z^2)^{k}$. One has to average over the unknown 
direction $\vec{n}$ to obtain
\be
p_k^{\mbox{\scriptsize sb}} = \int \frac{d^2\vec{n}}{4\pi} n_z^{2k} = \frac{1}{2k+1}.
\label{eq:sb}
\ee
One recovers in an easy calculation the large $N$ limit of the exact result,
Eq.(\ref{eq:exact_ratio})!
We note that the result (\ref{eq:sb}) is much larger than the naive (and wrong) result (\ref{eq:Naive}) 
as soon as $k\gg 1$.  In the limiting case where all the atoms of the gas have been detected, that is
$k=N$, we expect to see a difference between the symmetry-breaking prediction
and the exact result. The probability $p_N$ of finding all the $N$ spins
with vanishing angular momentum along $z$ is given for the exact ground
state by
\be
p_N = |\langle N:0_{\vec{e}_z}|\Psi_0\rangle|^2 = \frac{1}{N+1}
\ee
so that the symmetry breaking prediction $p_N^{\mbox{\scriptsize sb}}$ is too small by a factor 
of two. 

\section{The ground state of a 1D attractive Bose gas}
\label{Cap:soliton}
We consider in this section a model of a one-dimension Bose
gas with attractive interactions in the absence of confining potential.
This model is not so unrealistic at it may
appear at a first glance. Experimentally one can start with a three-dimensional Bose-Einstein condensate
with effective attractive interactions, that is with a negative scattering
length, as it is the case for lithium \cite{Hulet} or rubidium 85 \cite{JILA_85}.
The condensate is then
subject to a strong harmonic confinement in the $x-y$ plane, with a quantum of oscillation
along $x$ and $y$ larger than the absolute value of the
mean interaction energy per particle in the gas. In this way the motion of the condensate atoms
is frozen transversally in the ground state of the harmonic trap in the $x-y$ plane.
Along the $z$ direction one slowly reduces the trap strength so that the gas
becomes almost free.

Such a situation is interesting physically as it gives rise 
to the formation of \lq bright' solitons well
known in optics but not yet observed with atoms.
Moreover we have found exact $N-$body eigenstates for the
model of a one-dimension Bose gas with a $\delta$ attractive interaction
potential in free space. We use these exact solutions to test
the translational symmetry breaking Hartree-Fock approximation. 

\subsection{A model for the interaction potential}
By analogy with the three-dimensional model interaction potential
(\ref{eq:ppf}) we model the interaction in the one-dimension Bose
gas by a contact potential:
\be
V(z_1-z_2) = g_{1d} \delta(z_1-z_2).
\label{eq:pot1d}
\ee
The validity of such a modelization, and the value
of the coupling constant $g_{1d}$ in terms of the three-dimensional scattering
length $a$ are discussed in \cite{Maxim}.

The $\delta$ potential has the advantage in 1D of leading to a mathematically
well defined scattering problem, so that no regularization
operator is required, contrary to the 3D case.
In a practical solution of the $N-$body problem the use of a $\delta$
potential  amounts to
imposing some boundary conditions on an otherwise interaction free
Schr\"odinger equation \cite{Lieb,Gaudin}. 
Consider indeed an eigenstate of the $N-$body Hamiltonian with energy $E$:
\be
E \Psi(z_1,\ldots,z_N) =\left[ -\frac{\hbar^2}{2m} \sum_{i=1}^{N}
\partial_{z_i}^2  + g_{1d} \sum_{1\leq i < j \leq N}
\delta(z_i-z_j)\right] \Psi(z_1,\ldots,z_N).
\label{eq:schr}
\ee
The $N-$body wavefunction $\Psi$ is bosonic: it is symmetric
with respect to any permutation of the $N$ coordinates $z_1,\ldots,
z_N$ so it is sufficient to determine it on the fundamental domain
\be
D=\{(z_1,\ldots,z_N) \quad \mbox{such that}\quad z_1 < \ldots < z_N\}.
\ee
Inside the domain $D$ the positions of the particles are different so that
the $\delta$ terms can be omitted in (\ref{eq:schr}):
\be
\left[E+\frac{\hbar^2}{2m} \sum_{i=1}^{N} \partial_{z_i}^2\right]
\Psi =0
\label{eq:libre}
\ee
which is Schr\"odinger's equation for a non-interacting gas.
The $\delta$ terms are responsible for a discontinuity of the
spatial derivatives of $\Psi$ across the boundary of $D$. Using again
the bosonic symmetry of $\Psi$ one is able to relate a left derivative
(that is a derivative evaluated from the external side of $D$) 
to a right derivative (that is a derivative evaluated from the internal  side
of $D$) so that one obtains the conditions:
\be
\left(\partial_{z_{j+1}}-\partial_{z_j}\right )\Psi
\Big|_{z_{j+1}-z_j =0^+} 
= \frac{mg_{1d}}{\hbar^2}\Psi\Big|_{z_{j+1}=z_j}.
\label{eq:bc}
\ee

The differential equation (\ref{eq:libre}) over domain $D$,
the conditions (\ref{eq:bc}) on the boundary of $D$ and the condition
that $|\Psi|$ remains finite even for particles going to
$\pm\infty$ define the eigenstates of our model.

\subsection{Ground state of the one-dimension attractive Bose gas}
It turns out that in the previously exposed model 
one can calculate exactly eigenenergies and eigenstates of
the Hamiltonian for $N$ particles using the Bethe
ansatz \cite{Lieb,Gaudin} in the fundamental domain $D$:
\be
\Psi(z_1,\ldots,z_N) = {\cal N} \sum_{\sigma\in S_N}
A(\sigma) \exp\left(i\sum_{j=1}^{N} z_j k_{\sigma(j)}\right) \quad\mbox{for}\quad (z_1,\ldots,z_N)
\in D
\label{eq:ba}
\ee
where the $k_j$'s are real arbitrary wavevectors and where the sum is taken over
all permutations $\sigma$ 
of $N$ objects. This form is a superposition of plane waves in the domain $D$ and clearly solves
(\ref{eq:libre}) with the energy
\be
E=\frac{\hbar^2}{2m} \sum_{j=1}^{N} k_j^2.
\ee
The amplitudes $A(\sigma)$ of each permutation in the sum are
adjusted to satisfy the boundary conditions (\ref{eq:bc}):
\be
A(\sigma) = \prod_{1\leq j < l \leq N} \left(
1+\frac{img/\hbar^2}{k_{\sigma(j)} -k_{\sigma(l)}}
\right).
\label{eq:a_sigma}
\ee
In the case 
$g_{1d}>0$ one can show that the family of eigenstates obtained 
from the Bethe ansatz is complete \cite{Gaudin}. The proof of
completeness fails for the attractive case $g_{1d}<0$.

We are considering here the less studied attractive case $g_{1d}<0$.
Intuitively, the fact that this family of eigenstates is no
longer complete is not 
surprising as the Bethe ansatz leads to positive
eigenenergies only, whereas bound states with negative energies are expected to
exist for attractive interactions.
The way out is to extend the Bethe ansatz by adding an imaginary part $\kappa_j$
to the wavevectors:
\be
k_j \rightarrow K_j = k_j + i \kappa_j
\label{eq:subs}
\ee
where both $k_j$ and $\kappa_j$ are real.
After this substitution in the Bethe form (\ref{eq:ba}) the $N-$body wavefunction
still satisfies Eq.(\ref{eq:libre}) with an energy
\be
E= \frac{\hbar^2}{2m}\sum_{j=1}^{N} \left(k_j^2-\kappa_j^2 +2i\kappa_j k_j\right).
\label{eq:ener}
\ee
The boundary conditions (\ref{eq:bc}) are equally satisfied provided that
one performs the substitution (\ref{eq:subs}) in Eq.(\ref{eq:a_sigma}).
What remains to be verified is that the $N-$body wavefunction does
not diverge exponentially when one or several particles are going to
infinity! 
For the term of the generalized Bethe ansatz associated to the permutation
$\sigma$, this requires that either
\be
A(\sigma) =0
\label{eq:cond1}
\ee
or  the real part of the argument of the exponential in the Bethe
ansatz should not run off to $+\infty$ over the domain $D$
\cite{proviso}:
\be
\sum_{j=1}^{N} z_j \kappa_{\sigma(j)} > -\infty\quad\mbox{over}\quad D.
\label{eq:cond2}
\ee
The second condition (\ref{eq:cond2}) can be rewritten in
a more explicit way, by taking $z_1$ (varying from $-\infty$ to $+\infty$)
and $Z_2=z_2-z_1, \ldots, Z_N=z_N-z_{N-1}$ (varying from $0$ to $+\infty$)
as independent variables. Equation (\ref{eq:cond2}) is then equivalent to the set of conditions
\bea
\sum_{j=1}^{N} \kappa_j &=& 0
\label{eq:cond2_1} \\
\sum_{j=l}^{N} \kappa_{\sigma(j)} &\geq & 0 \quad\mbox{for}\quad l=2,\ldots,N.
\label{eq:cond2_2}
\eea
Once these conditions are satisfied one can hope to have a vanishing imaginary
part of the energy $E$ in Eq.(\ref{eq:ener})!

We have found a general form for the $K_j$ satisfying for all $\sigma$'s one of the two
conditions (\ref{eq:cond1},\ref{eq:cond2}).
We discuss here the set of $K_j$ corresponding to the
absolute ground state of the gas, the case of excited states is considered
in \S\ref{subsec:excited}. First it is found that all the real parts
$k_j$ are vanishing, which indeed is an efficient way to minimize the energy
$E$ as the $k_j$'s contribute as squares with a positive coefficient!
Second the imaginary parts $\kappa_j$ are given by
\be
\kappa_j = \frac{m|g_{1d}|}{2\hbar^2} \left[2j -(N+1)\right]
\quad \mbox{for}\quad j=1,\ldots,N.
\label{eq:choice}
\ee
The corresponding ground state energy was already known \cite{Guire}:
\be
E_0(N) = -\frac{1}{24} \frac{m g_{1d}^2}{\hbar^2} N (N^2-1).
\label{eq:exe}
\ee
Let us check that the choice (\ref{eq:choice}) indeed leads to a $N-$body wavefunction
not diverging at infinity.

A first way to proceed is to show that conditions (\ref{eq:cond1}) or (\ref{eq:cond2})
are satisfied. First one remarks that a permutation $\sigma$ different from
identity will have a vanishing coefficient $A(\sigma)$. The reciprocal $\sigma^{-1}$
of such a permutation cannot
indeed conserve the numerical ascending order of the integers $1,\ldots,N$ so that
there exists an integer $n$ satisfying
\be
\sigma^{-1}(n+1) < \sigma^{-1}(n).
\ee
If one considers in $A(\sigma)$ the factor of indices $j<l$ with
\be
j=\sigma^{-1}(n+1) \quad\mbox{and}\quad  l=\sigma^{-1}(n)
\ee
one gets a vanishing factor in Eq.(\ref{eq:a_sigma}):
\be
1+\frac{i m g_{1d}/\hbar^2}{i(\kappa_{n+1}-\kappa_n)} =0
\ee
according to Eq.(\ref{eq:choice}), leading to $A(\sigma)=0$.
Second one can check by direct substitution that conditions (\ref{eq:cond2_1},\ref{eq:cond2_2})
are satisfied, using the identity
\be
\sum_{j=l}^{N}\left[ 2j-(N+1)\right] = (N+1-l)(l-1).
\ee

A second way to show that the proposed ground state wavefunction cannot blow up at infinity
is to realize that, up to a normalization factor ${\cal N}$,
it can be rewritten as
\be
\Psi_0(z_1,\ldots,z_N) = {\cal N}
\exp\left[\frac{m g_{1d}}{2\hbar^2}\sum_{1\leq i < j\leq N} |z_i-z_j|\right],
\label{eq:psi_ex}
\ee
for {\bf all} values of the coordinates $z_j$'s, not restricting to the
fundamental domain $D$.
This form, already known in the literature, 
coincides with the generalized Bethe ansatz over the fundamental 
domain $D$ and is obviously symmetric with respect to any permutation
of particles. As the argument inside the exponential in (\ref{eq:psi_ex})
is negative, the expression (\ref{eq:psi_ex}) is clearly bounded from above.
To determine the normalization factor $\cal N$ we enclose the gas in a 
fictitious box of size $L$ tending to $+\infty$; this leads to \cite{math}:
\be
|{\cal N}|^2 = \frac{(N-1)!}{NL}\left(\frac{m|g_{1d}|}{\hbar^2}\right)^{N-1}.
\ee

\subsection{Hartree-Fock approximation for the ground state}
To what extent can we recover the results of the previous
subsection using a Hartree-Fock ansatz
$|N:\phi\rangle$ for the ground state wavefunction? We calculate the mean energy
of the Hartree-Fock ansatz, and we obtain the following energy functional of
$\phi$, sum of kinetic energy and mean field interaction energy:
\be
E^{\mbox{\scriptsize hf}}[\phi,\phi^*]= 
N \int dz \left[ {\hbar^2\over 2m}\left|\frac{d\phi}{dz}\right|^2 +
{1\over 2}(N-1)g_{1d}|\phi(z)|^4\right].
\ee
This functional has to be minimized over $\phi$ with the constraint
that $\phi$ is normalized to unity. A pure dimensional analysis gives already the main
feature of the result. Let us rescale the coordinate $z$ with some length $\xi$:
\be
\phi(z) = \frac{1}{\sqrt{\xi}} \psi(x=z/\xi)
\ee
such that the kinetic energy term and the mean field energy term have equal
coefficients:
\be
\frac{\hbar^2}{m\xi^2} = \frac{(N-1)|g_{1d}|}{\xi}
\label{eq:xi}
\ee
so that the energy functional becomes
\bea
E^{\mbox{\scriptsize hf}}[\phi,\phi^*] &=& \frac{m g_{1d}^2}{\hbar^2} N(N-1)^2 \epsilon[\psi,\psi^*] \\
\epsilon[\psi,\psi^*] &=& \int dx\, \left[\frac{1}{2}|\psi'(x)|^2 -\frac{1}{2}|\psi(x)|^4\right].
\eea
We have already obtained the minimal Hartree-Fock energy up to a numerical factor.

The minimization of the energy functional $\epsilon$ gives access to this numerical factor.
One can restrict to a real wavefunction $\psi(x)$ since a $x-$dependent phase for a fixed
modulus immediately increases the kinetic energy without decreasing the interaction
energy. The fact that the functional $\epsilon$ is stationary around the minimal energy
solution$\psi$ leads to the adimensioned Gross-Pitaevskii equation:
\be
\frac{1}{2}\psi''(x) - \psi^3(x) = \nu \psi(x)
\label{eq:gpe}
\ee
where the adimensioned chemical potential $\nu$ is a Lagrange multiplier
ensuring that $\psi$ remains normalized to one in the variation. After multiplication
of (\ref{eq:gpe})
by $\psi'(x)$ one can integrate over $x$ to obtain:
\be
-\frac{1}{4}\psi'^2(x) -\frac{1}{4}\psi^4(x) -\frac{1}{2}\nu \psi^2(x)=0
\ee
where the integration constant has been taken equal to zero as $\psi$ vanishes
at $x=\pm\infty$. Note that the quantity $\nu$ is then clearly negative.
The problem is reduced to a quadrature and one finds:
\be
\psi(x) = \frac{\sqrt{2|\nu|}}{\cosh[\sqrt{2|\nu|}(x-x_0)]}
\ee
where $x_0$ is arbitrary.
As the integral over all the real axis $y$ of $1/\cosh^2(y)$ is equal to two, we obtain
\be
\nu= -\frac{1}{8}.
\ee
We have recovered the well-known solitonic solution for the one-dimension non-linear
Schr\"o\-din\-ger equation:
\be
\phi_{z_0}(z) = \frac{1}{2\xi^{1/2}} \frac{1}{\cosh[(z-z_0)/(2\xi)]}
\label{eq:soli}
\ee
parameterized by the arbitrary position $z_0$ of the soliton.
Redimensioning the quantity $\nu$ gives 
the chemical potential of the soliton \cite{remarque}:
\be
\mu =  - \frac{1}{8} \frac{m g_{1d}^2}{\hbar^2} (N-1)^2.
\label{eq:mu}
\ee
The final result for the Hartree-Fock minimal energy is of course independent
of $z_0$:
\be
E_0^{\mbox{\scriptsize hf}}(N) = -\frac{1}{24} \frac{m g_{1d}^2}{\hbar^2} N (N-1)^2.
\label{eq:hfe}
\ee

The deviation of the Hartree-Fock minimal energy (\ref{eq:hfe}) 
from the exact result (\ref{eq:exe}) is a fraction $1/N$
of the energy and is small in the large $N$ limit. Can we understand why
the validity condition of the Hartree-Fock ansatz is simply $N\gg 1$?
The result is not intuitive as large values of $N$ lead to small spatial
widths $\xi$ of the soliton, and therefore to high linear densities of
particles, where one may expect to have a strongly interacting
regime not well described by mean field. The paradox can be removed
in the following way:
the Hartree-Fock ansatz for the $N-$body wavefunction does not contain any
correlation between the particles, as it is a factorized state vector, 
so it is an acceptable approximation only if the interaction potential is weak
enough, that is if the interaction potential can be treated in the Born
approximation for the relevant relative momenta of the particles.
An exact calculation of the scattering amplitude of two particles 
with initial relative wavevector $k$ and interacting with the model
potential (\ref{eq:pot1d}) shows that the Born approximation
is applicable provided that $k$ is high enough:
\be
\left|\frac{\hbar^2 k}{m g_{1d}} \right| \gg 1
\label{eq:ni}
\ee
in contrast to the three-dimensional case. Taking the estimate $k\simeq 1/\xi$
for the typical relative momentum of particles we find that condition
(\ref{eq:ni}) reduces to $N\gg 1$. Actually the fact that the weakly
interacting regime in one-dimension corresponds to a {\bf high} density regime
of the gas is a well established fact also for effective repulsive
interactions $g_{1d}>0$ \cite{Lieb}.

There is however a notable difference of translational properties of the exact
$N-$body ground state (\ref{eq:psi_ex}) and of the Hartree-Fock ansatz.
Whereas the exact ground state is invariant by a global translation
of the positions of the particles, as it should be, the Hartree-Fock ansatz
leads to condensate wavefunctions $\phi_{z_0}$ localized within the length $\xi$ around some
arbitrary point $z_0$, see Eq.(\ref{eq:soli}).
The Hartree-Fock ansatz
$|N:\phi_{z_0}\rangle$ therefore breaks the translational symmetry of the system.

Breaking a symmetry of the system costs energy, and this can be checked
for the present translational symmetry breaking.
As the center of mass coordinate and momentum $Z,P$ of the $N$
particles are decoupled from the relative coordinates of the particles
we can write the total energy of the gas as the
sum of the kinetic energy of the center of mass 
and of an \lq internal' energy including
the kinetic energy of the relative motion of the particles and the interaction
energy. 
Whereas the exact ground state wavefunction has a vanishing
center of mass kinetic energy, the symmetry breaking state $|N:\phi_{z_0}\rangle$
contains a center of mass kinetic energy:
\be
E_{\mbox{\scriptsize c.o.m.}}^{\mbox{\scriptsize hf}} =  
\langle N:\phi_{z_0}|\frac{\hat{P}^2}{2mN}| N:\phi_{z_0}\rangle
\ee
where $mN$ is the total mass of the gas and $\hat{P}=p_1+\ldots+p_N$ is the 
total momentum operator of the gas. Expanding 
the square of $\hat{P}$,
and using the fact that the soliton wavefunction $\phi$ has 
a vanishing mean momentum we obtain
\bea
E_{\mbox{\scriptsize c.o.m.}}^{\mbox{\scriptsize hf}} &=& \langle \phi_{z_0} |\frac{p^2}{2m}|\phi_{z_0}\rangle \\
&=& \frac{1}{24} \frac{mg_{1d}^2}{\hbar^2}(N-1)^2.
\label{eq:ekin}
\eea
We see that $E_{\mbox{\scriptsize c.o.m.}}^{\mbox{\scriptsize hf}}$ accounts for half the energy
difference between the exact ground state energy (\ref{eq:exe})
and the Hartree-Fock energy (\ref{eq:hfe}). 

The restored symmetry ansatz of Eq.(\ref{eq:ps}):
\be
|\Psi_0\rangle^{\rm hfs} =  \lim_{L\rightarrow +\infty}\frac{\cal N}{L^{1/2}}
\int_{-L/2}^{L/2} dz_0 |N:\phi_{z_0}\rangle
\ee
has the advantage of having
a vanishing total momentum and will have a lower energy than the symmetry
breaking ansatz. We can try to calculate its energy. First we have to
evaluate overlap integrals of $\phi_{z_0}$ and $\phi_{z_0'}$, with $z_0\neq z_0'$:
these integrals can be brought in the form of integrals of rational functions
using a change of variable $z=2\xi \exp u$; these integrals over $u$ can be calculated
exactly. We are left with an integral
over $z_0=x/(2\xi)$:
\bea
E_0^{\rm hfs}(N) &=& E_0^{\rm hf}(N)
\frac{\displaystyle\int_{-\infty}^{+\infty} dx\, A(x) \left(\frac{x}{\sinh x}\right)^N}
{\displaystyle\int_{-\infty}^{+\infty} dx  \left(\frac{x}{\sinh x}\right)^N} \\
A(x) &=& \frac{3}{2\sinh^2 x}\left(3+\cosh 2x -\frac{2\sinh 2x}{x}\right)
+\frac{6}{x\sinh x}\left(\cosh x -\frac{\sinh x}{x}\right).
\eea
This ratio of integrals can be evaluated in the large $N$ limit, using Laplace's method.
The function $x/\sinh(x)$ raised to the large power $N$ is very peaked around $x=0$
with a width scaling as $1/\sqrt{N}$ 
so that one can expand in powers of $x=y/\sqrt{N}$:
\bea
\left(\frac{x}{\sinh x}\right)^N &=& \exp\left[-\frac{1}{6}y^2+\frac{1}{180}\frac{y^4}{N} + O(N^{-2})
\right] \\
A(x) &=& 1 + \frac{2}{5} \frac{y^2}{N} + O(N^{-2}).
\eea
We are left with Gaussian integrals that can be performed exactly. We compare the
various results to the exact ground state energy:
\bea
E_0^{\rm hf}(N) &=& E_0(N) \left[ 1-\frac{2}{N} + O(N^{-2})\right] \\
E_0^{\rm hfs}(N) &=& E_0(N) \left[ 1-\frac{4}{5N} + O(N^{-2})\right].
\eea
We see that the symmetrized Hartree-Fock prescription (\ref{eq:ps}) has a lower
energy than the symmetry breaking prescription (\ref{eq:do}) but the relative
error
is of the same order $O(1/N)$.

\subsection{Physical advantage of the symmetry breaking description}
We now raise the question: is there a Bose-Einstein condensate in the one-dimension
unconfined Bose gas with attractive interaction? To make things simple we assume that the
gas is at zero temperature so that the $N-$particle wavefunction is known
exactly, see Eq.(\ref{eq:psi_ex}).

We start with a reasoning in terms of the one-body density operator (even if we know
from the previous physical example that this may be dangerous, see \S\ref{subsec:advan}). 
Paraphrasing the
usual definition of a Bose-Einstein condensate in 
three dimensional free space
we put the one-dimension gas in a fictitious box of size $L$ 
and we calculate the mean number of particles $n_0$ in the plane wave with vanishing momentum
$p=0$ in the limit $L\rightarrow +\infty$.

The calculation with the exact ground state wavefunction can be done, it
is too involved to be reproduced here.
One finds that $n_0$ is going to zero as $1/L$:
\be
n_0 \simeq \frac{C(N)}{L} \frac{2\hbar^2}{m|g_{1d}|}.
\label{eq:exact_n0}
\ee
The factor $C(N)$ is given by 
\be
C(N) = \sum_{i=1}^{N}\sum_{j=i}^{N}
\frac{(j-1)!}{(i-1)!}\frac{(N-i)!}{(N-j)!}
\prod_{k=i}^{j}\left[k(N+1-k)-\frac{1}{2}(N+1)\right]^{-1}
\ee
and converges to $\pi^2/2$ in the large $N$ limit.
There can be therefore no macroscopic population in the $p=0$
momentum state in the large $L$ limit.
One may then be tempted to conclude 
that there is no Bose-Einstein condensate,
even at zero temperature, in the free one-dimension Bose gas with attractive contact
interactions. However we have learned that a reasoning based on the one-body
density matrix may miss crucial correlations between the particles, and
that the symmetry breaking point of view may be illuminating in this
respect.

The translational symmetry breaking point of view approximates the state of the gas by
the $N-$body density operator:
\be
\hat{\rho}^{\mbox{\scriptsize sb}} = \lim_{L\rightarrow +\infty}
\int_{-L/2}^{L/2} \frac{dz_0}{L} |N:\phi_{z_0}\rangle \langle N:\phi_{z_0}|.
\label{eq:sba}
\ee
In the large $N$ limit we expect this prescription to be valid
for few-body observables. Of course for a
$N-$body observable such as the kinetic energy of the center of mass of the gas,
the results will be different: the kinetic energy
Eq.(\ref{eq:ekin}) for the symmetry-breaking
point of view differs from the exact vanishing value.

Let us test this expectation by calculating in the Hartree-Fock 
approximation 
the mean 
number of particles in the plane wave $\langle z|k\rangle=\exp(ikz)/L^{1/2}$.
Using the following action of the annihilation operator $\hat{a}_k$ of a particle
with wavevector $k$ on the Hartree-Fock state:
\be
\hat{a}_k |N:\phi_{z_0}\rangle = N^{1/2} \langle k|\phi_{z_0}\rangle
|N-1:\phi_{z_0}\rangle
\ee
we obtain
\be
n_k^{\mbox{\scriptsize hf}}  = N |\langle k|\phi_{z_0}\rangle|^2.
\label{eq:mom_dist}
\ee
The momentum distribution of the particles in the gas in this approximation
is simply proportional to the momentum distribution of a single particle
in the solitonic wavefunction $\phi_{z_0}$! 
It turns out that the Fourier transform of the $1/\cosh$ function
can be calculated exactly, and it is also a $1/\cosh$ function.
We finally obtain:
\be
n_k^{\mbox{\scriptsize hf}}  \simeq \frac{1}{L} \frac{\pi^2\hbar^2}{m |g_{1d}|}
\frac{1}{\cosh^{2}\left(\pi k \xi\right)}
\label{eq:nks}
\ee
where $\xi$ is the typical soliton size given in Eq.(\ref{eq:xi}).
For $k=0$ one recovers 
the large $N$ limit of the exact result (\ref{eq:exact_n0}).

In more physical terms, one can imagine from Eq.(\ref{eq:sba}) that 
a given experimental realization of the Bose gas corresponds to  
a condensate of $N$ particles in the solitonic wavefunction
(\ref{eq:soli}), with a central position $z_0$ being a random variable varying
in an unpredictable way for any new realization of the experiment.  
There is therefore
a Bose-Einstein condensate in the one-dimension attractive Bose gas!

An illustrative gedanken experiment would be to measure the positions along $z$
of all the particles of the gas.
In the symmetry breaking point of view the positions $z_1,\ldots, z_N$ obtained in
a single measurement are randomly distributed according to the density $|\phi_{z_0}^2|
(z)=|\phi_0(z-z_0)|^2$
where $z_0$ varies from shot to shot as the relative
phase of the two condensates did in the MIT interference experiment. 
As we know the exact ground
state (\ref{eq:psi_ex}) we also know the exact $N-$body distribution function,
$|\Psi_0(z_1,\ldots,z_N)|^2$. This is however not so easy to use! 

So we suggest instead to consider 
the mean spatial density of the particles {\sl knowing} that the center
of mass of the cloud has a position $Z$. In the exact formalism this gives after
lengthy calculations:
\bea
\rho(z|Z) &=& \int dz_1\ldots\int dz_N |\Psi_0(z_1,\ldots,z_N)|^2
\left(\sum_{j=1}^{N}\delta(z-z_j)\right)
L\delta\left(Z-\frac{1}{N}\sum_{n=1}^{N}z_n\right)
\label{eq:def_rho}\\
&=&   \frac{ N}{\xi} \sum_{k=0}^{N-2}
\frac{(N-2)!}{(N-k-2)!}  \frac{N!}{(N+k)!}(-1)^k(k+1)
\exp\left[-(k+1)\frac{N}{N-1}\frac{|z-Z|}{\xi}\right]
\nonumber
\eea
where $\xi$ is the $N-$dependent length of the soliton (\ref{eq:xi}),
the integrals are taken in the range $[-L/2,L/2]$ and $L\rightarrow
+\infty$; the factor $L$, compensating the one in the normalization
factor of $\Psi$, ensures that the integral of $\rho(z|Z)$
over $z$ is equal to $N$.

In the symmetry breaking point of view the definition of $\rho(z|Z)$ 
is similar to Eq.(\ref{eq:def_rho}); 
the factor $L$ cancels with the $1/L$ factor of Eq.(\ref{eq:sba}).
This leads to
\bea
\rho^{\mbox{\scriptsize sb}}(z|Z) &=& \int\! dz_0 \int\! dz_1\ldots\int\! dz_N \left(\prod_{k=1}^{N}
|\phi_{z_0}(z_k)|^2\right)
\left(\sum_{j=1}^{N}\delta(z-z_j)\right)
\delta\left(Z-\frac{1}{N}\sum_{n=1}^{N}z_n\right)
\nonumber
\\
&=& N\int dz_1\ldots\int dz_N \left(\prod_{k=1}^{N}|\phi_0(z_k)|^2\right)
\delta\left(Z-z+z_1-\frac{1}{N}\sum_{n=1}^{N}z_n\right)
\label{eq:complique}
\eea
where we have made the change of variables $z_k\rightarrow z_k+z_0$ (which allows
to integrate over $z_0$), we recall that $\phi_0$ is the solitonic wavefunction
centered in $z_0=0$ and we have replaced the sum over the indistinguishable particles
$j$ by $N$ times the contribution of particle $j=1$.
The multiple integral over the positions $z_1,\ldots,z_N$
can be turned into a single integral over a wavevector $q$ by using the identity
$\delta(X)= \int dq/(2\pi) \exp(iqX)$, allowing a numerical calculation of
$\rho^{\mbox{\scriptsize sb}}(z|Z)$. 

\begin{figure}[htb]
\centerline{ 
\epsfysize=7cm \epsfbox{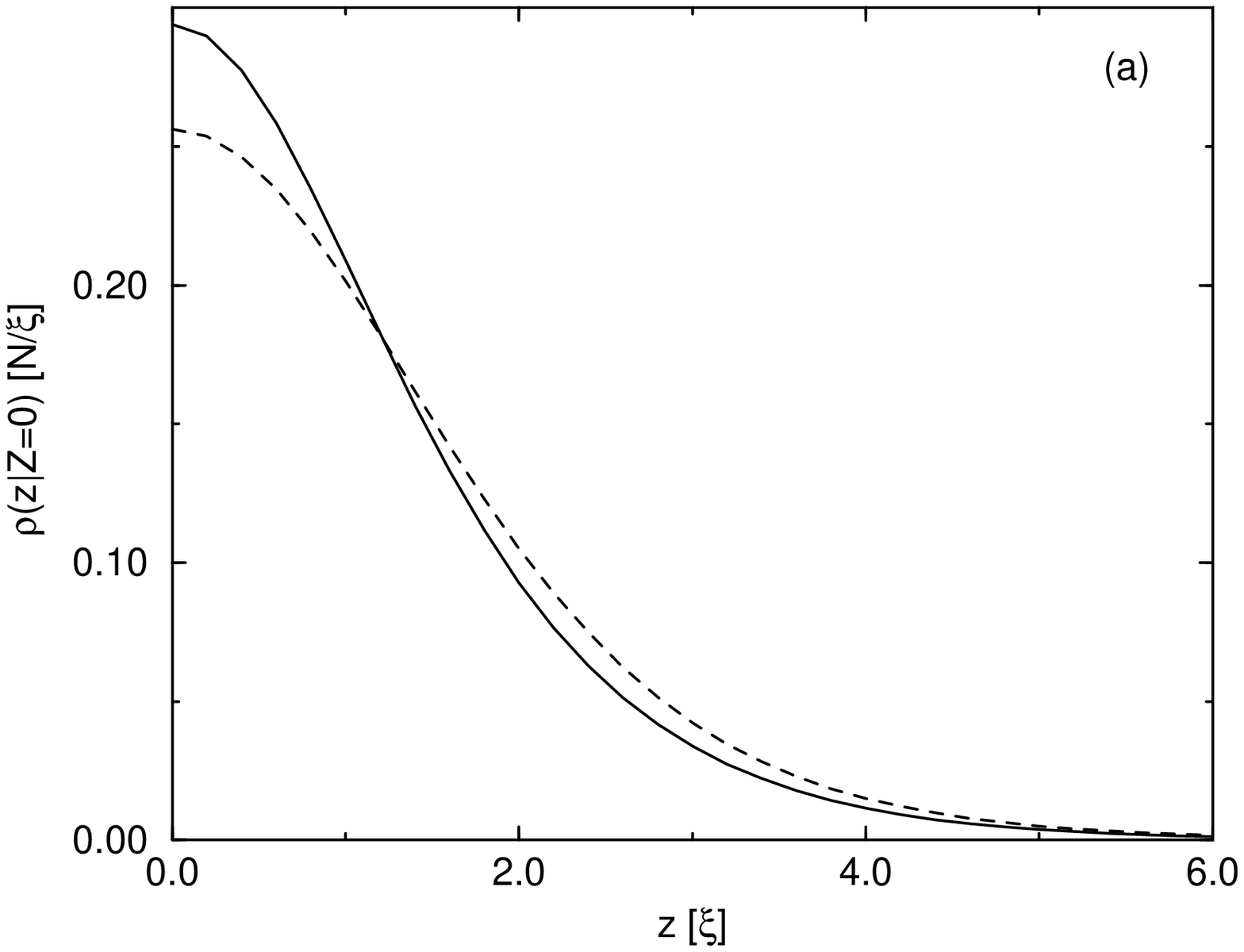} \ \ \
\epsfysize=7cm \epsfbox{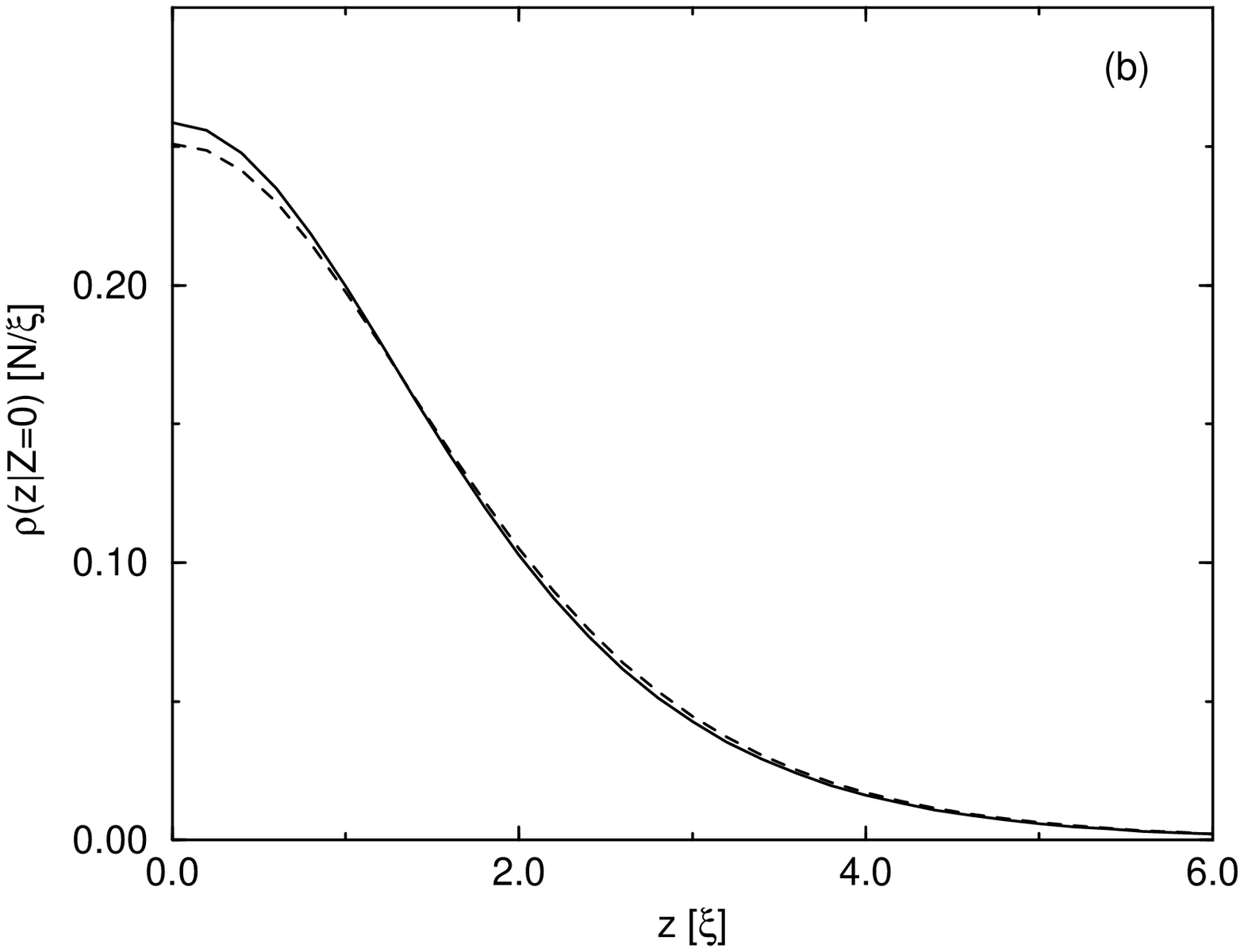}}
\caption{\small For the ground state of the one-dimension attractive Bose gas,
position dependence of the
mean density of particles knowing that the center-of-mass of the gas is
in $Z=0$. Solid line: exact result $\rho(z|Z=0)$. Dashed line: mean-field approximation
$\rho^{\mbox{\scriptsize sb}}(z|Z=0)$. The position $z$ is expressed in units
of the \lq soliton' radius $\xi$ where $\xi$ is given in Eq.(\ref{eq:xi}), and the linear
density in units
of $N/\xi$. The number of particles is (a) $N=10$ and (b) $N=45$. 
\label{fig:soli}}
\end{figure}

Does the approximate result (\ref{eq:complique}) 
get close to the exact result
for large $N$?  We compare numerically in figure~\ref{fig:soli} 
the exact density $\rho(z|Z)$ to the symmetry breaking mean-field prediction
$\rho^{\mbox{\scriptsize sb}}(z|Z)$:
modestly large values of $N$ give already good agreement between the two densities.
This validates the symmetry breaking approach for the considered gedanken
experiment.

What happens in the large $N$ limit?
In Eq.(\ref{eq:complique}) each variable $z_k$ explores an interval of size
$\sim \xi$ so that the quantity $(z_1+\ldots+z_N)/N$ has a standard
deviation $\sim \xi/\sqrt{N}$ much smaller than $\xi$ and can be neglected as compared
to $z_1$ inside the $\delta$ distribution. This leads to
\be
\rho^{\mbox{\scriptsize sb}}(z|Z) 
\simeq N|\phi_{z_0=Z}(z)|^2\ \ \ \mbox{for}\ \ \ \sqrt{N}\gg 1.
\label{eq:proper}
\ee
Numerical calculation of $\rho^{\mbox{\scriptsize sb}}(z|Z)$ 
shows that Eq.(\ref{eq:proper})
is a good approximation over the range $|z-Z|\simeq \xi$ for $N=10$
already!

\subsection{Excited states of the one-dimension Bose gas}\label{subsec:excited}
We briefly show how to calculate excited states of the gas which can contain 
clusters or \lq lumps' of bound particles, and that are therefore not contained in the usual
Bethe ansatz (\ref{eq:ba}). These excited states are obtained with the generalized Bethe
ansatz involving complex wavevectors $K_j$ as defined in Eq.(\ref{eq:subs}).

The first step is to group the atoms in an arbitrary number $R$ of lumps ($1\leq R\leq N$), 
the lump number $r$
containing $n_r$ particles. More precisely, the positions of the particles are in the
fundamental domain $D$, that is they are in ascending order $z_1 < \ldots < z_N$; we put
the particles $1,\ldots,n_1$ in the first lump, the particles $n_1+1,\ldots,n_1+n_2$ in
the second lump, etc, the particles $N-n_R+1,\ldots, N$ in the last, $R^{\mbox{\scriptsize th}}$
lump. It is then more convenient to reindex the particles: each atom is now identified
by its lump number $r$ (from $1$ to $R$) and its rank inside the lump $i_r$ (from $1$ to $n_r$)

\begin{figure}[htb]
\setlength{\unitlength}{0.5in}
\begin{center}
\begin{picture}(10,6)
\put(0,3){\vector(1,0){10}}
\put(5,0){\vector(0,1){6}}
\put(1,3.5){\circle*{0.2}}
\put(1,2.5){\circle*{0.2}}
\put(2.7,4){\circle*{0.2}}
\put(2.7,5){\circle*{0.2}}
\put(2.7,1){\circle*{0.2}}
\put(2.7,3){\circle*{0.2}}
\put(2.7,2){\circle*{0.2}}
\put(4.1,4.5){\circle*{0.2}}
\put(4.1,3.5){\circle*{0.2}}
\put(4.1,1.5){\circle*{0.2}}
\put(4.1,2.5){\circle*{0.2}}
\put(5.7,3){\circle*{0.2}}
\put(6.5,3){\circle*{0.2}}
\put(8,4){\circle*{0.2}}
\put(8,5){\circle*{0.2}}
\put(8,6){\circle*{0.2}}
\put(8,3){\circle*{0.2}}
\put(8,2){\circle*{0.2}}
\put(8,1){\circle*{0.2}}
\put(8,0){\circle*{0.2}}
\put(9.5,3){\circle*{0.2}}
\put(10,2.5){$k$}
\put(4.5,5.75){$\kappa$}
\end{picture}
\end{center}
\caption{
We plot the momenta $K_j=k_j + i\kappa_j$ for an arbitrary
excited state of $N=21$ particles with 7 lumps.
}
\label{fig:lump}
\end{figure}
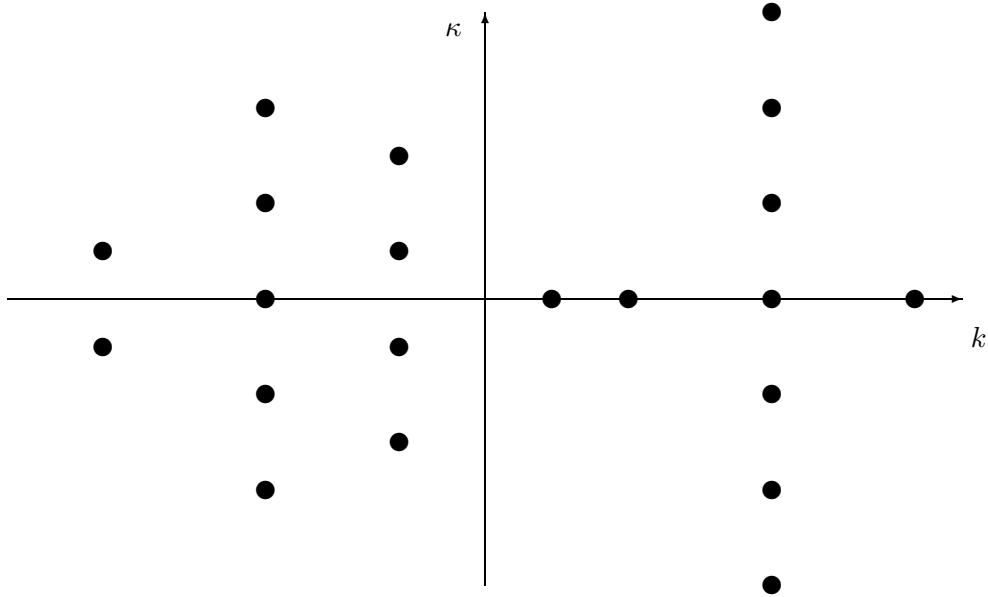

The complex wavevectors are then taken as
\be
K_{r,i_r} = k_r + i \frac{m |g_{1d}|}{2\hbar^2} [2 i_r -(n_r+1)].
\label{eq:choice2}
\ee
as examplified in figure~\ref{fig:lump}.
This choice has a simple physical interpretation. The imaginary parts of the $K_j$'s
inside lump $r$ with $n_r$ particles
are exactly the $\kappa_j$'s defining the ground state of a gas of
$n_r$ atoms. The real part of the $K_j$'s inside lump $r$ is the same for the $n_r$
particles of the lump and corresponds to a global motion of the lump with a 
total lump momentum of $n_r\hbar k_r$: this is nothing but an excitation of the center of mass
motion of the particles in the considered lump. The corresponding $N-$body wavefunction
is then simply a coherent ensemble of $R$ quantum solitons with different
momenta $\hbar k_r$.
The eigenenergy associated to the choice (\ref{eq:choice2}) can be calculated
from formula (\ref{eq:ener}):
\be
E = \sum_{r=1}^{R} \left(E_0(n_r) + n_r \frac{\hbar^2 k_r^2}{2m}\right)
\label{eq:exs}
\ee
where the ground state energy $E_0$ is given in (\ref{eq:exe}) as function of the number
of particles. We therefore find that the $R$ quantum solitons do not interact: this
is the quantum analog of the known fact that classical field solitons can cross 
without interaction.

What remains to be checked is that the proposed $N-$body wavefunction does not
explode at infinity. The proof proceeds along the same lines as for the ground state.
One first shows that only the permutations $\sigma$ such that
$\sigma^{-1}$ conserves the
ascending order of particle labels inside each lump can have a non-vanishing
factor $A(\sigma)$. That is $\sigma$ has to satisfy
\be
\sigma^{-1}(r,1)< \ldots <\sigma^{-1}(r,i_r)<\ldots < \sigma^{-1}(r,n_r) \quad
\mbox{for all}\quad r=1,\ldots,R.
\ee
Then the conditions (\ref{eq:cond2_1},\ref{eq:cond2_2}) are checked to be
satisfied by direct substitution.

It is interesting to compare the exact spectrum (\ref{eq:exs}) to mean field
predictions. We restrict here \cite{full_spectrum} to a simple calculation using the Bogoliubov
approach to calculate the excitation spectrum of the classical soliton. This amounts
to linearizing the time dependent Gross-Pitaevskii equation
\be
i\hbar \partial_t \phi(z,t) = -\frac{\hbar^2}{2m}\partial_z^2 \phi(z,t)+g_{1d}(N-1) |\phi(z,t)|^2
\phi(z,t) -\mu \phi(z,t)
\ee
around the stationary solution (\ref{eq:soli}) and to looking for corresponding eigenmodes.
One finds that the eigenmode $(u(z),v(z))$ with energy $\epsilon$ solves the usual
Bogoliubov-de Gennes equations:
\bea
\epsilon u(z) &=& \left[-\frac{\hbar^2}{2m} \frac{d^2}{dz^2} + 2g_{1d} (N-1) |\phi_{z_0}|^2 -\mu\right] u(z) +
g (N-1) \phi_{z_0}^2 v(z) 
\label{eq:u}\\
-\epsilon v(z) &=& \left[-\frac{\hbar^2}{2m}\frac{d^2}{dz^2} + 2g_{1d} (N-1) |\phi_{z_0}|^2 -\mu\right] v(z) +
g (N-1) \phi_{z_0}^{*2} u(z).
\label{eq:v}
\eea
The Bogoliubov modes have long been known in the context of optical solitons
\cite{Kaup}. Apart from two zero energy modes corresponding to the $U(1)$ symmetry
and translational symmetry Goldstone modes, the eigenmodes behave as traveling waves
$\exp(i k z)$ far away from the center of the soliton. The eigenenergy is straightforward 
to obtain: far from the soliton we can set $\phi_{z_0}\simeq 0$ in (\ref{eq:u}) and we get
the dispersion relation
\be
\epsilon_k  = \frac{\hbar^2 k^2}{2m} -\mu
\ee
corresponding to a gap $|\mu|$ in the excitation spectrum.
In the large $N$ limit the Hartree-Fock chemical potential $\mu$ of (\ref{eq:mu}) 
is very close to the exact chemical potential $E_0(N)-E_0(N-1)$. In this way
the elementary excitation energy $\epsilon_k$ is very close to the energy
difference between the absolute ground state of the gas and the excited state
with two lumps, one lump with $N-1$ particles at rest and the other lump
with one particle with momentum $\hbar k$! The Bogoliubov approach is however unable
to predict bound states between excitations, corresponding to excited lumps
with two or more than two particles.

\section{Conclusion}
We have studied in great details two exactly solvable
models of the interacting Bose gas. The first model involves $N$ interacting 
spin one particles assumed to be in the ground state of
a cubic box, the second model is a one-dimension Bose gas of spinless particles
interacting with an attractive contact potential in the absence of external 
confinement.

Both models share the property of not leading to the formation
of a Bose-Einstein condensate, even at zero temperature,
in the strict Penrose and Onsager sense: the one-body density operator 
has three macroscopic eigenvalues for the spins, and has no macroscopic
eigenvalue for the attractive Bose gas.

We have shown that the usual Hartree-Fock approach, trying to
approximate the ground state of the gas by a Hartree-Fock state 
with all the $N$ particles in the same single particle state $\phi$,
can bring a considerable physical understanding of these model
systems. A continuous family of $\phi$'s is found to minimize the
mean energy, each member of the family having the same energy 
and corresponding to a state with broken rotational symmetry (for the spins)
or translational
symmetry (for the one-dimension gas). 
This allows to imagine that any particular experimental
realization of the gas is a condensate in the broken symmetry state $\phi$,
with $\phi$ varying in an unpredictable way 
for any new realization of the experiment. E.g.\ one imagines that 
any particular realization of the one-dimension attractive Bose
gas is a condensate in a soliton wavefunction.

We have successfully tested the Hartree-Fock approximation against
the exact results, in the thermodynamical limit for the spins
and in the large $N$ limit for the one-dimension Bose gas. 
In particular the prediction of the issues of gedanken
experiments performed on the gas was found to be much more transparent
in the symmetry-breaking point of view than with the exact $N-$body
ground state wavefunction.

An interesting aspect has not been discussed in the present article.
It involves the compared robustness of the exact ground state
and the Hartree-Fock states against decoherence induced {\sl e.g}.\ by particle
losses due to three body collisions \cite{Sinatra}. 
While the Hartree-Fock states $|N:\phi\rangle$
remain Hartree-Fock states under removal of one or several particles and are
therefore robust, the exact ground state, after having experienced
several particle losses, becomes a density operator that may actually
look like a statistical mixture of Hartree-Fock states,
these losses being formally equivalent to
measurements performed on the system \cite{mesure_franges2}.
Such an aspect is important
in the context of an experimental attempt to distinguish 
between the exact and the symmetry breaking descriptions.

We acknowledge useful discussions with Gordon Baym, Jean Dalibard, Anthony Leggett,
Maxim Olshanii and Sorin Paraoanu.
We are grateful to Alice Sinatra for a critical reading of the manuscript. 


\renewcommand{\baselinestretch} {1}


\begin{thebibliography}{99}
\bibitem{Haar} One has to use the so-called Haar invariant measure for the integration
over the symmetry group.
\bibitem{proviso}
As the complex wavevectors $K_j$ are all distinct (in order for the denominators
in the $A(\sigma)$'s not to vanish) the various terms of the generalized Bethe ansatz
with $A(\sigma)\neq 0$
are functionally linearly independent: none of the terms
in the generalized Bethe ansatz should diverge at infinity over the domain $D$.
\bibitem{math}
The center of mass of the gas corresponds to a fictitious particle of wavevector
$K$, where $\hbar K$ is the total momentum of the gas, and of position $Z$,
where $Z$ is the centroid of the gas. In the ground state $|\Psi\rangle$
the center of mass is completely delocalized with $K=0$. The
factor $1/L$ in $|{\cal N}|^2$ originates
from the normalization of the fictitious
particle plane wave in the fictitious box of size $L$,
$\langle Z |K\rangle = e^{i K Z}/\sqrt{L}$.
The more correct mathematical way (not used here) is to normalize in free
space (no box) using the closure relation
$\int dK |K\rangle\langle K| = \mbox{Id}$, which amounts to replace
$L$ by $2\pi$.
\bibitem{remarque} We do not have here exactly $\mu = \partial_N E_0^{\mbox{\scriptsize hf}}$
as we have not approximated $N-1$ by $N$ in the energy functional,
so $\mu$ is more precisely the \lq eigenvalue' of the Gross-Pitaevskii equation
satisfied by $\phi$. The difference is however negligible is the large $N$ limit.
\end{thebibliography}
\end{document}